%% file: main.tex
\documentclass[10pt, conference]{IEEEtran}

\usepackage[square,numbers,sort&compress]{natbib}
\usepackage{tabularx, color, colortbl}
\usepackage{graphicx, subfigure}
\usepackage{amsmath, amssymb}
\usepackage{algorithm}
\usepackage{algpseudocode}
\usepackage{listings}
\usepackage[noblocks]{authblk}
\usepackage{mathtools}
\usepackage{todonotes}
\usepackage{framed}
\usepackage{subfig}
\usepackage{soul}
\usepackage{verbatim}
\usepackage{url}
\usepackage{booktabs}
\usepackage{flushend}
\usepackage{comment}
\usepackage{multirow}
\usepackage{subfiles}
\usepackage{booktabs}
\usepackage{caption}
\usepackage{subcaption}
\usepackage{siunitx}

\usepackage[letterpaper, margin=0.75in]{geometry}

\usepackage{cleveref}
\crefname{section}{§}{§§}
\Crefname{section}{§}{§§}

\begin{document}



\title{IOAgent: Democratizing Trustworthy HPC I/O Performance Diagnosis Capability via LLMs}

\author[1]{Chris Egersdoerfer}
\author[2]{Arnav Sareen}
\author[3]{Jean Luca Bez}
\author[4]{Suren Byna}
\author[5]{Dongkuan (DK) Xu}
\author[1]{Dong Dai}

\affil[1]{University of Delaware, USA, 
\{cegersdo, dai\}@udel.edu}

\affil[2]{University of North Carolina at Charlotte, USA, 
asareen2@charlotte.edu}

\affil[3]{Lawrence Berkeley National Laboratory, USA,  jlbez@lbl.gov}

\affil[4]{The Ohio State University, USA,
byna.1@osu.edu}

\affil[5]{North Carolina State University, USA,
dxu27@ncsu.edu}
\maketitle

\begingroup
\renewcommand\thefootnote{}
\footnotetext{Published in the Proceedings of the 2025 IEEE International Parallel and Distributed Processing Symposium (IPDPS 2025).}
\endgroup


\input{sections/0_abs}
\input{sections/1_intro}
\input{sections/2_background}
\input{sections/3_pre_study}
\input{sections/4_design}
\input{sections/5_dataset}
\input{sections/6_experiments}
\input{sections/7_conclusion}

\section*{Acknowledgments}
We sincerely thank the anonymous reviewers for their valuable feedback. This work was supported in part by National Science Foundation (NSF) under grants CNS-2008265, CCF-2412345, OAC-2417850, and BCS-2416846. This effort was also supported in part by the U.S. Department of Energy (DOE), Office of Science, Office of Advanced Scientific Computing Research (ASCR) under contract number DE-AC02-05CH11231 with LBNL and under an LBNL subcontract to OSU (GR130493).

\bibliographystyle{plain}
\bibliography{references}
\end{document}

%% file: sections/0_abs.tex
\begin{abstract}
As the complexity of the HPC storage stack rapidly grows, domain scientists face increasing challenges in effectively utilizing HPC storage systems to achieve their desired I/O performance. To identify and address I/O issues, scientists largely rely on I/O experts to analyze their I/O traces and provide insights into potential problems. However, with a limited number of I/O experts and the growing demand for data-intensive applications, inaccessibility has become a major bottleneck, hindering scientists from maximizing their productivity. 

The recent rapid progress in large language models (LLMs) opens the door to creating an automated tool that democratizes trustworthy I/O performance diagnosis capabilities to domain scientists. However, LLMs face significant challenges in this task, such as the inability to handle long context windows, a lack of accurate domain knowledge about HPC I/O, and the generation of hallucinations during complex interactions. 

In this work, we propose \texttt{IOAgent} as a systematic effort to address these challenges. IOAgent integrates various new designs, including a module-based pre-processor, a RAG-based domain knowledge integrator, and a tree-based merger to accurately diagnose I/O issues from a given Darshan trace file. Similar to an I/O expert, IOAgent provides detailed justifications and references for its diagnoses and offers an interactive interface for scientists to continue asking questions about the diagnosis. To evaluate IOAgent, we collected a diverse set of labeled job traces and released the first open diagnosis test suite, \texttt{TraceBench}. Based on this test suite, extensive evaluations were conducted, demonstrating that IOAgent matches or outperforms state-of-the-art I/O diagnosis tools with accurate and useful diagnosis results. We also show that IOAgent is not tied to specific LLMs, performing similarly well with both proprietary and open-source LLMs. We believe IOAgent has the potential to become a powerful tool for scientists navigating complex HPC I/O subsystems in the future.
\end{abstract}

%% file: sections/1_intro.tex
\section{Introduction}
\label{sec:intro}
High-performance computing (HPC) clusters play a crucial role in enabling modern science, supporting a wide range of scientific applications and services across various domains, such as cosmology~\cite{almgren2013nyx}, quantum simulation~\cite{kim2018qmcpack}, astronomy~\cite{jacob2009montage}, climate modeling~\cite{kurth2018exascale}, and life sciences~\cite{1000genome-workflow}.

In recent years, these scientific applications have become increasingly data-intensive, necessitating the efficient utilization of HPC I/O subsystems to meet performance requirements. However, effectively leveraging the complex I/O stacks in HPC systems, which include high-level parallel I/O libraries (e.g., HDF5~\cite{HDF5}, PnetCDF~\cite{PnetCDF}), API interfaces (e.g., MPI-IO, POSIX, STDIO), and file systems (e.g., Lustre~\cite{Lustre}, BurstBuffer~\cite{henseler2016architecture}), remains a challenging task for most domain scientists. Large-scale scientific applications often experience slow I/O performance due to inefficient use of data access APIs, misconfigurations, or the failure to apply key optimizations available in the storage systems~\cite{tavakoli2016log,rashid2023iopathtune}.

One proven approach to assist scientists in optimizing their I/O performance is to record the I/O traces of their applications for post-hoc analysis~\cite{zhang2023drill, egersdoerfer2022clusterlog}. Real-time I/O profiling tools, such as Darshan~\cite{darshan} and Recorder~\cite{Recorder}, have been developed and deployed in modern HPC facilities to collect detailed I/O traces. These traces provide comprehensive information about the I/O behavior of applications, including access patterns, I/O sizes, and timing data to help scientists understand their application's I/O behaviors. Additional tools, such as PyDarshan~\cite{pyDarshan} and DXT-Explorer~\cite{dxt-explorer}, have been introduced to interpret these traces, identify potential I/O issues, and even offer optimization suggestions~\cite{egersdoerfer2023early, zhang2021sentilog}.

Despite these advancements, diagnosing I/O issues from application traces often requires the involvement of human I/O experts, who possess the specialized knowledge needed to interpret trace data due to the inherent complexity of modern HPC storage systems. With the increasing number of scientists from various domains developing HPC applications, the shortage of readily available I/O expertise presents a significant barrier. Our interviews with I/O experts from NERSC also confirm that a substantial backlog of applications with I/O performance issues is frequently awaiting analysis, with no readily available means to absolve such complications.

This gap is further exacerbated by the growing complexity of HPC systems and the vast amounts of data generated by modern applications. As a result, the difficulty in identifying and resolving I/O performance bottlenecks limits the potential of scientists and leads to inefficient use of computational resources. Therefore, there is an urgent need for an automated tool that can democratize access to I/O optimization expertise.

Recent developments in large language models (LLMs) like ChatGPT and Claude have shown some promises in addressing such challenges. Pre-trained on large datasets, these models demonstrate ingenious abilities in understanding and generating human-like text, making them highly accessible. Their in-context learning ability~\cite{salewski2023incontext}, combined with prompt engineering techniques~\cite{brown2020languagemodelsfewshotlearners} and Chain-of-Thought (CoT) reasoning~\cite{wei2023chainofthoughtpromptingelicitsreasoning}, enables them to follow instructions and incorporate new information. Real-world examples span areas such as medical questioning~\cite{liévin2023large}, military simulation and strategy~\cite{rivera2024escalation}, log-based anomaly detection~\cite{egersdoerfer2023early}, and educational guidance~\cite{xiao-etal-2023-evaluating}. Furthermore, recent advancements in Retrieval-Augmented Generation (RAG) allow LLMs to integrate external information from knowledge bases during the generation process~\cite{lewis2020retrieval}, significantly enhancing their ability to produce accurate or contextually appropriate responses by incorporating information beyond their pre-training. 
The capability of continuous interactions between LLM and users is another unique capability, making it extremely useful as an assistant in many tasks.


These distinctive and compelling capabilities of LLMs illuminate a path to democratize domain scientists’ access to HPC I/O optimization. Given an I/O trace, LLMs may automate the I/O performance diagnostic process, making HPC I/O optimization expertise more accessible to domain scientists and removing the necessity for a human expert. This can ultimately accelerate scientific discovery and enhance computational efficiency.

However, realizing such an ambitious goal is not straightforward, with traditional LLMs facing multiple significant challenges that inhibit their ability to interpret, analyze, and process I/O traces.
First, although traces such as those generated by Darshan can be parsed into a human-readable format and hence are directly interpreted by LLMs, their lengths often surpass millions of lines which exceeds typical LLMs' context window size. Due to this, directly querying LLMs with such traces leads to a multitude of problems and in many cases, an inability to leverage the model entirely. 
For example, LLMs are known to encounter issues with long context windows, such as \textit{lost-in-the-middle truncation}, where LLMs truncate the input, resulting in much of the information contained within center of the context being ignored in favor of text at extremities of the document~\cite{lostinmiddle}, and \textit{losing long-range dependencies}, where LLMs struggle to maintain coherence across distant parts of the input, preventing them from effectively modeling dependencies across lengthy documents~\cite{borgeaud2022improvinglanguagemodelsretrieving}. 
Unfortunately, in practice, critical information about I/O issues could be located anywhere in the trace, such as inefficient I/O behavior for a subset of files in the middle of the application execution. Additionally, many I/O issues can only be identified by correlating multiple parts of the I/O trace, such as the amount of MPI-IO vs. the amount of POSIX IO, making it challenging, even for state-of-the-art LLMs to accurately diagnose I/O problems, as exemplified in the next section.

Second, diagnosing I/O performance issues from I/O traces requires extensive domain knowledge, as these issues may be deeply rooted in specific I/O libraries or embedded within the nuances of storage systems. However, such domain-specific knowledge might not be available to LLMs during their training stage, especially considering that new findings are continuously being published. Even if domain knowledge is available to LLMs during training, it may not be effectively utilized due to the vast corpus that these models are exposed to, which typically encompasses an extensive amount of general information, thus diluting the specificity necessary to effectively tackle domain-specific tasks (as demonstrated in an example later). While this generality is typically a prominent strength of modern LLMs, it significantly limits its capability to act with expert-level proficiency in a particular domain.

Third, LLMs are known to produce hallucinations, plausible but factually incorrect or non-sensical outputs, which are unacceptable for use as a diagnosis tool by domain scientists~\cite{ji2023survey}. Although opting for more advanced models such as GPT-4o instead of simpler versions such as GPT-4 or open-source models like Llama~\cite{touvron2023llamaopenefficientfoundation}, can help reduce hallucinations~\cite{zhao2024wildhallucinations}, larger models consequently introduce higher costs~\cite{OpenAIpricing}.
As a tool targeting general scientists working at the HPC scale, relying on expensive models is impractical. Therefore, addressing the hallucination challenge independently of an individual model is crucial to enable an LLM-based tool at the extensive magnitude commonly present in HPC applications.

In this study, we explore the feasibility and strategy of using large language models to provide trustworthy, expert-level HPC I/O performance diagnosis, thereby effectively guiding domain scientists in optimizing their I/O performance.
To this end, we introduce \texttt{IOAgent}, an LLM-based agent designed to analyze applications’ I/O traces. IOAgent incorporates key new designs to address the three challenges discussed earlier. First, IOAgent implements a module-based trace preprocessing strategy that groups relevant I/O activities together, preventing LLMs from truncating vital context or missing key I/O modules. Additionally, IOAgent introduces a set of summarization methods that accurately extract needed metadata from the trace file, rather than relying solely on the limited capabilities of LLMs for metadata retrieval.
Second, IOAgent utilizes Retrieval-Augmented Generation (RAG) to integrate expert-level I/O knowledge into the diagnosis process. This allows IOAgent to provide references for its diagnoses, helping to avoid popular but incorrect claims while also increasing the transparency of the diagnosis process.
Third, IOAgent employs a tree-based merging mechanism with self-reflection to minimize hallucinations in its diagnosis. 
These mechanisms enable IOAgent to provide domain scientists with accurate diagnostic summaries of I/O issues based on Darshan I/O traces, akin to consulting with a human expert, even if the scientists utilize open-source LLMs such as Llama.

To evaluate IOAgent and similar tools that may be developed in the future, we created a new benchmark suite for I/O issue diagnosis, named \texttt{TraceBench}. TraceBench includes over {$40$} Darshan traces collected from both I/O benchmarks and real-world applications, with each trace annotated with issues identified by I/O experts, and each I/O issue confirmed by at least two experts. By evaluating the diagnostic results with these standard labels, one can assess and compare the quality of diagnostic tools from a variety of sources. Using \texttt{Tracebench}, we demonstrated that IOAgent outperforms state-of-the-art I/O diagnostic tools, such as Drishti~\cite{drishti} and ION~\cite{egersdoerfer2024ion} in terms of accuracy, clarity, and coverage.

The core value of IOAgent lies not only in diagnosing HPC I/O performance issues but also in its ability to provide valuable design choices on how to leverage powerful but difficult-to-manage LLMs for production-level systems, where both accuracy and cost are of utmost importance. We believe that similar tools can be developed to democratize other optimization capabilities in the HPC environment, such as computation and networking in the near future.


%% file: sections/2_background.tex
\section{Background and Related Work}
\label{sec:background}

\begin{figure*}[h!]
    \centering
\includegraphics[width=0.8\textwidth]{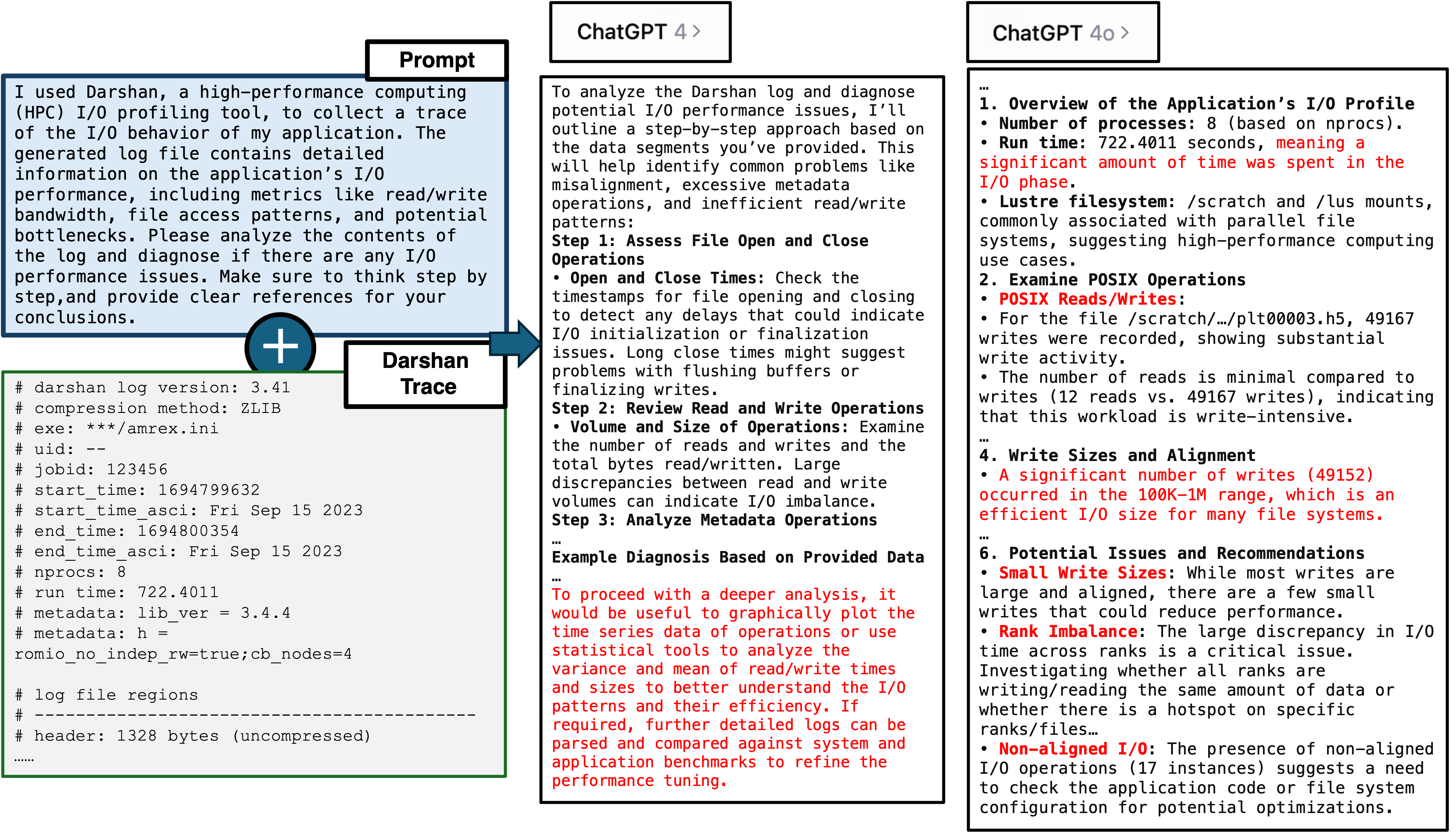}
    \caption{An example diagnosis done by directly querying two LLMs: \textit{gpt-4} and \textit{gpt-4o}.}
    \label{llm-example}
\end{figure*}

In this section, we provide an overview of HPC I/O profiling tools, focusing on Darshan and its extensions. We also discuss how researchers have utilized these tools to understand HPC I/O characteristics and its typical {workflow}. Subsequently, we introduce Drishti as a state-of-the-art I/O issue diagnosis tool. Finally, we provide background knowledge on large language models, their reasoning capabilities, and the concept of Retrieval-Augmented Generation (RAG).

\subsection{Darshan: HPC I/O Profiling Tool}
Efficient I/O performance is critical for HPC applications, and as a result, understanding I/O behavior is essential for optimization. Multiple I/O profiling tools have been developed to facilitate this understanding, including STAT~\cite{STAT}, mpiP~\cite{mpip}, IOPin~\cite{IOPIN}, Recorder~\cite{Recorder}, and Darshan~\cite{darshan}. Among these, Darshan has gained widespread adoption in the HPC community due to its lightweight design and holistic descriptions of applications~\cite{moti2023trace}.

Darshan operates by instrumenting HPC applications to record key statistical metrics related to their I/O activities. In particular, it traces essential information for each file accessed across different I/O interfaces, including POSIX (Portable Operating System Interface) I/O, MPI (Message Passing Interface) I/O, and Standard I/O. The metrics collected encompass multiple aspects and can be mainly summarized as follows: \texttt{1) data volume}, amount of data read from and written to each file, 2) \texttt{operation counts}, number of read, write, and metadata operations performed by the application, 3) \texttt{temporal information}, aggregate time spent on read, write and metadata operations, 4) \texttt{rank information}, identification of the MPI ranks issuing I/O requests, and 5) \texttt{variability metrics}, variance of I/O sizes and timing among different application ranks.

Darshan also collects file system-specific metrics, such as Lustre stripe widths and Object Storage Target (OST) IDs over which a file is striped. This comprehensive data provides valuable insights into the I/O patterns and performance characteristics of HPC applications.

Darshan eXtended Tracing (DXT)~\cite{dxt} is a recent development based on Darshan to capture fine-grained records of an application’s I/O operations, covering details such as specific files involved in I/O operations, each read/write operation, offset, and length of each I/O request, the start and end times of each I/O operation, and MPI rank IDs. Researchers have utilized Darshan DXT to analyze I/O behaviors across a variety of applications and systems \cite{patel2020gift}. However, since Darshan DXT introduces more noticeable overheads to HPC applications, it is typically not enabled by default. Hence, in this study, we focus only on the original Darshan I/O traces and leave working with Darshan DXT traces as future work. 



\subsection{Advanced I/O Issue Diagnosis: Drishti and ION}
While tools like Darshan provide the raw data necessary for I/O analysis, interpreting this data to diagnose performance issues remains challenging. Several tools have been developed to address this gap, including IOMiner~\cite{IOMiner}, UMAMI~\cite{UMAMI}, TOKIO~\cite{TOKIO}, DXT-Explorer~\cite{dxt-explorer}, recorder-viz~\cite{Recorder}, and Drishti~\cite{drishti}. Of these, Drishti represents the most recent advancement in I/O issue diagnosis.

Drishti takes a Darshan trace as input and conducts an analysis to report various I/O performance issues based on a set of heuristic-based triggers. Currently, Drishti includes a set of 30 triggers corresponding to different application behaviors and can identify nine distinct types of I/O issues, such as \texttt{small I/O operations} (excessively small read/write requests), \texttt{misaligned I/O} (I/Os not aligned with the file system’s block size), and \texttt{imbalanced I/O} (uneven distribution of I/O workload across ranks).

Due to its lightweight design of checking key triggers in Darshan traces, Drishti can be an effective tool for system administrators or scientists to quickly scan a large number of traces and identify applications with I/O issues. However, it has notable limitations when it comes to providing detailed I/O diagnoses for domain scientists working on their applications. 
First, Drishti relies on hard-coded threshold values (determined via expert knowledge and observations from past experience) for its triggers, which may not be accurate for all applications. For instance, it flags write requests smaller than 1MB as small writes and raises an issue if more than 10\% of all requests are small I/Os. While 10\% is a reasonable threshold for general applications, this can be perplexing to domain scientists if their application only exhibits a minute percentage of such operations, with the resulting performance degradation being negligible. Domain scientists would benefit far more from a tool that provides personalized, per-application explanations for each diagnosis.
Second, Drishti's explanations and recommendations are pre-defined, hard-coded messages tied to their specific triggers. This approach lacks the nuanced and context-specific reasoning needed for domain scientists to fully understand and address performance problems. 
Lastly, Drishti does not offer an interactive interface for users to ask follow-up questions or further explore the analysis. This reduces its utility as a learning and exploratory tool, especially for domain scientists who may not have extensive I/O expertise.

These limitations motivated ION~\cite{egersdoerfer2024ion}, a recent study that also uses large language models to diagnose HPC I/O issues. As a proof-of-concept work, ION directly queries the LLMs with engineered prompts to generate diagnoses. This strategy, however, means ION will heavily rely on the capability of the selected LLMs as well as suffer from their shortcomings, which will be further discussed later in Section~\ref{sec:llm}. 

\subsection{Large Language Models and Their Capabilities}

Recent advancements in artificial intelligence have led to the development of large language models (LLMs) like ChatGPT, Claude AI, and Gemini~\cite{gemini}, which demonstrate remarkable capabilities in understanding and generating humanesque text. 
One of the key strengths of LLMs is their ability to follow instructions and address various tasks using their \textit{in-context learning} capability. They can perform complex tasks such as summarization~\cite{zhang2023benchmarkinglargelanguagemodels}, translation~\cite{brown2020languagemodelsfewshotlearners}, and question answering~\cite{chang2023surveyevaluationlargelanguage}, and even exhibit emergent abilities like mathematics and programming reasoning~\cite{wei2022emergentabilitieslargelanguage}. This makes them well-suited for applications that require comprehension and synthesis of information across diverse domains.

Another significant strength of LLMs is their enhancement via the utilization of \textit{Retrieval-Augmented Generation (RAG)}~\cite{lewis2020retrieval}. RAG combines LLMs with external knowledge bases or documents, enabling the model to retrieve relevant information during the generation process. This approach enhances the model’s ability to provide accurate and up-to-date information, particularly in specialized domains where the model’s pre-training data may be insufficient and advancements are made rapidly.

In the context of HPC I/O analysis, LLMs offer a promising opportunity to automate the diagnostic process. However, applying LLMs to interpret I/O trace data and pinpoint I/O issues requires addressing several critical challenges, as discussed earlier. 
Recognizing both the upside potential and challenges of applying LLMs to HPC I/O analysis, our goal in this study is to develop a solution that harnesses the strengths of LLMs while mitigating their limitations.


%% file: sections/3_pre_study.tex
\section{Preliminary Study: Plain LLM Diagnosis}
\label{sec:llm}

\begin{figure*}[h!]
    \centering
\includegraphics[width=\textwidth]{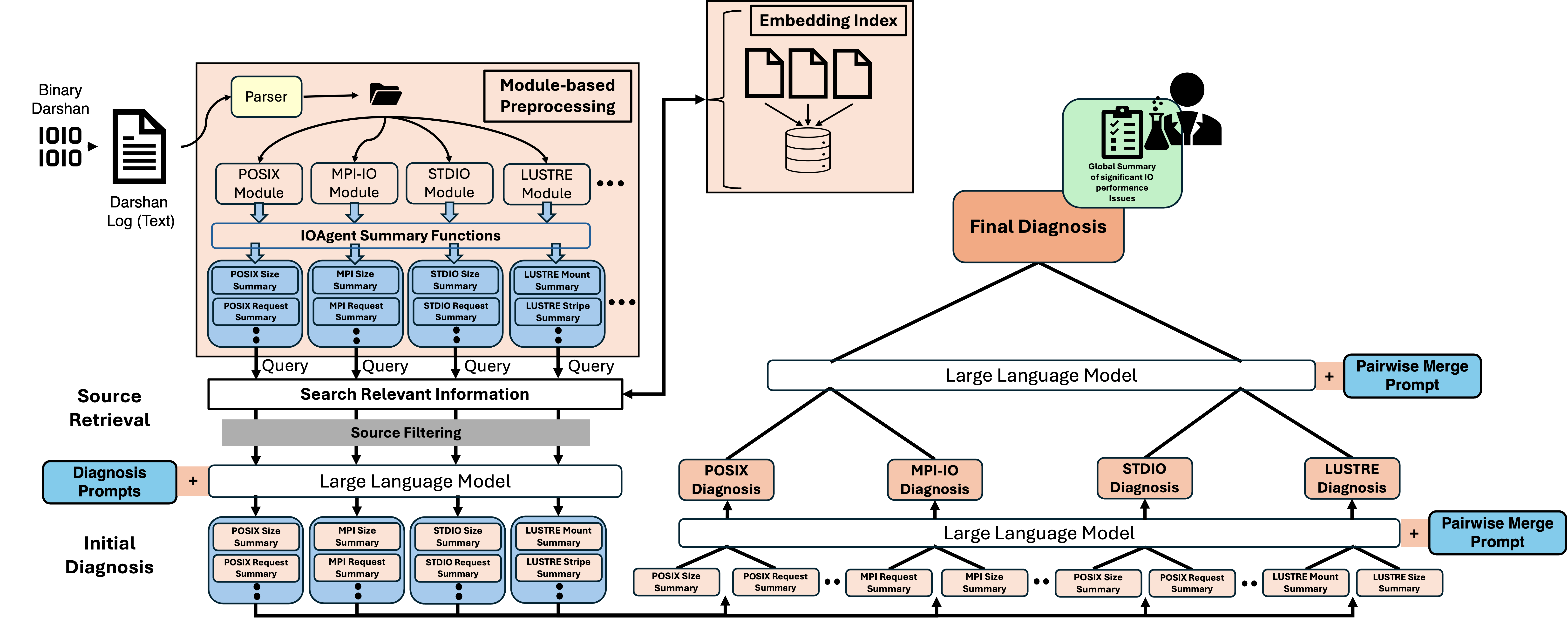}
    \caption{The overall workflow of IOAgent and its key components.}
    \label{ioagent}
    \vspace{-1em}
\end{figure*}

In this section, we first report the diagnostic results of a real-world HPC application's Darshan trace using plain language models. For this, we collected the Darshan trace of AMReX running at the NERSC HPC center. AMReX~\cite{zhang2021amrex} is a widely used framework for highly concurrent, block-structured adaptive mesh refinement. This AMReX execution took around 722 seconds to finish, used 8 processes, and read/write 11 files to a Lustre file system mounted at \texttt{/scratch}. 

Since the original Darshan trace is in binary format, we first used \texttt{darshan-parser} to convert it into a text-based, human-readable format. We then queried various language models using the prompt shown in Figure~\ref{llm-example}. Due to space constraints, we omit the outputs of open-source models such as Llama~\cite{touvron2023llamaopenefficientfoundation}, as the quality of their results is significantly lower compared to OpenAI's GPT models.

The outputs from the ChatGPT-4 model are not particularly useful. It primarily generates a simple plan to assess different aspects of the I/O traces, such as open/close operations, read/write operations, metadata operations, and stripe patterns, among others. In the end, it provides high-level recommendations on what domain scientists should do for deeper analysis, which are hardly useful; they are still left with the burden of learning how to `\textit{graphically plot the time series data of operations or use statistical tools…}' to uncover potential I/O issues.

Compared to ChatGPT-4, the new 4o model makes significant progress in addressing the I/O diagnosis problem, as shown by the output on the right of Figure~\ref{llm-example}. It follows the prompt’s instructions accurately, checking the I/O details and providing corresponding diagnoses. Most of the diagnoses are accompanied by concrete data about the I/O operations, such as `\textit{The file alignment was set at 1MB (1048576 bytes), which matches the common Lustre stripe size. This is optimal for minimizing the number of I/O requests on Lustre,}'. 

Although promising, ChatGPT-4o’s outputs actually contain multiple incorrect answers, an issue commonly seen with plain LLM-based agent tools~\cite{huang2023surveyhallucinationlargelanguage}. 
First, due to context truncation, it misses or forgets important information in the I/O traces. In this AMReX run, a key I/O performance issue is the predominant use of the POSIX interface for I/O instead of MPI-IO, which is expected to offer better performance at this scale. The 4o model overlooks this issue, likely because the information about the MPI-IO model appears in the latter half of the Darshan trace.
Second, the 4o model tends to overly rely on widely prevalent misconceptions that are believed but not necessarily true due to the broad-scoped nature of how LLMs are trained, which can lead to incorrect diagnoses. For example, it outputs “\textit{…1MB…stripe size…is optimal for minimizing the number of I/O requests on Lustre},” but given that the \texttt{stripe count} is set to $1$ in this application, such a configuration actually restricts parallel I/O capabilities, resulting in limited bandwidth and parallelism. Additionally, the “small write sizes” issue reported by the 4o model at the end may confuse domain scientists, as it previously reports “\textit{a significant number of writes (49152) occurred in the 100K-1M range, which is an efficient I/O size}.” This inconsistency in the diagnosis makes it difficult for domain scientists to fully trust or apply the tool.

We could not present the results from the latest OpenAI \texttt{o1-preview} model due to its limited context window, which is too small to process the complete AMReX Darshan trace. Our evaluations using smaller Darshan traces show that \texttt{o1-preview} produces outputs of similar quality to the 4o model. However, the high cost of \texttt{o1-preview} makes it largely impractical for our large-scale use.

These issues highlight the limitations of naively applying large language models to the I/O performance diagnosis task. We propose IOAgent, an LLM-based I/O diagnosis tool designed with key features to overcome these deficiencies.

%% file: sections/4_design.tex
\section{IOAgent Design and Implementation}
\label{sec:design}

In order to address the aforementioned challenges which non-augmented LLMs face when analyzing I/O trace logs, IOAgent mainly consists of three primary components. The first component, the \textit{module-based pre-processor} primarily handles the context-length challenge faced by LLMs and guides IOAgent towards effective downstream source retrieval. The second component, the \textit{Domain Knowledge Integrator} alleviates the non-augmented LLM's gaps in specific domain expertise with up-to-date and relevant information. The final component, the \textit{Tree-Merge} (with self-reflection), enables complex summarization for both frontier and short-of-frontier language models to form comprehensive and interpretable I/O performance diagnoses.

\begin{table*}[ht!]
\centering
\begin{tabular}{lcccccccccc}
\toprule
\textbf{Module} & \multicolumn{10}{c}{\textbf{Summary Category}} \\ 
 \midrule \midrule
 & \textbf{I/O Size} & \textbf{I/O Request Count} & \textbf{File} & \textbf{Metadata} & \textbf{Rank} & \textbf{Alignment} & \textbf{Order} & \textbf{Mount} & \textbf{Stripe Setting} & \textbf{Server Usage} \\
\hline
POSIX  & \textbf{\checkmark} & \textbf{\checkmark} & \textbf{\checkmark} & \textbf{\checkmark} & \textbf{\checkmark} & \textbf{\checkmark} & \textbf{\checkmark} & - & - & - \\ \hline
MPIIO  & \textbf{\checkmark} & \textbf{\checkmark} & \textbf{\checkmark} & \textbf{\checkmark} & \textbf{\checkmark} & - & - & - & - & - \\ \hline
STDIO  & \textbf{\checkmark} & \textbf{\checkmark} & \textbf{\checkmark} & - & - & - & - & - & - & - \\ \hline
LUSTRE & - & - & - & - & - & - & - & \textbf{\checkmark} & \textbf{\checkmark} & \textbf{\checkmark} \\ \hline
\end{tabular}
\caption{Coverage of Summary Categories Across Darshan Modules}
\label{Darshan_coverage}
\vspace{-1em}
\end{table*}

\subsection{Module-based Pre-processor}

To address the challenge of integrating HPC trace logs into the limited context window of LLMs, IOAgent uses a log pre-processor based on two key insights. First, effective and comprehensive reasoning over a trace log requires access to data from all key modules used by the application. To meet this need, IOAgent's pre-processor separates the Darshan log into a set of CSV files, with each file containing the counters and values from a single Darshan module. 

Second, since module data may also be extensive, it must be reduced to a manageable length for the LLM to interpret. IOAgent accomplishes this through summary extraction functions for each module, which generate categorized summary fragments. The categories of these summaries are outlined by the column titles in Table \ref{Darshan_coverage}. Due to the variance in counters accumulated by Darshan for each supported module, each module extracts varying categories of summary information. Due to these deviations, each module’s summary category uses its own extraction function based on the available counters, but the information follows consistent principles across categories. For example, the I/O volume function for the STDIO module extracts the total amount of read and written bytes, while for LUSTRE, it focuses on information about mount points, stripe settings, and server usage.

\begin{figure}[t!]
    \centering
    \includegraphics[width=0.45\textwidth]{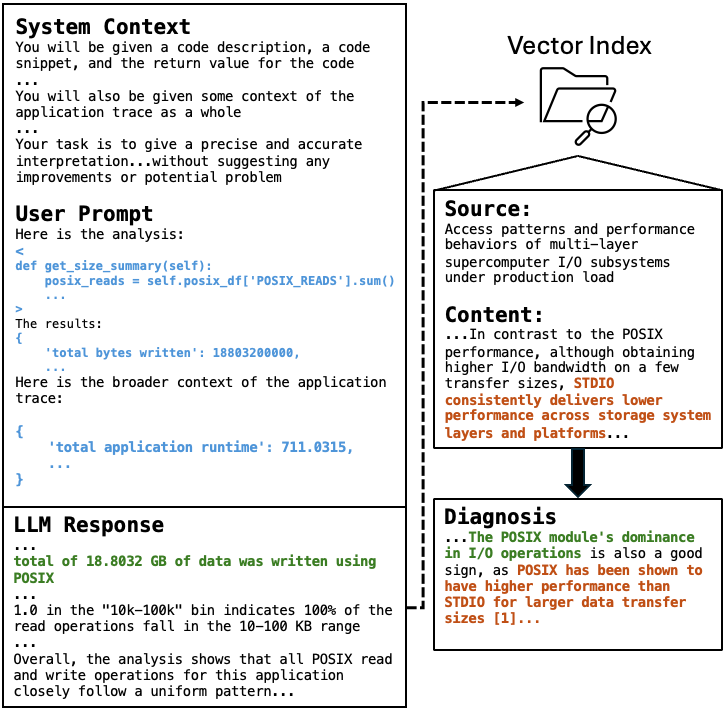}
    \caption{An example prompt and response transforming JSON summary fragment to natural language}
    \label{pre_rag}
    \vspace{-1em}
\end{figure}

\subsection{Domain Knowledge Integrator}
IOAgent uses Retrieval-Augmented Generation (RAG) to integrate domain knowledge to provide trustworthy diagnosis for HPC I/O issues. There are three key questions while using RAG: 1) how to construct the queries, 2) how to build the vector database, 3) and what are the key vector search parameters. We describe each of them in detail below.

\subsubsection{RAG Query Construction}
RAG provides a database for the LLM to query and retrieve useful and relevant information for their current task. Hence, the quality of the query to the RAG database matters significantly to retrieve the needed information.

After module-based pre-processing, we split module information into CSV files and compress it into manageable JSON summary fragments, which helps avoid overwhelming the limited context of LLMs. Each JSON fragment represents specific information from a Darshan module, making it easier to locate relevant information using techniques like cosine similarity between text embeddings. However, JSON summaries are still not in the same format as expert knowledge, which is often communicated through research papers. This makes direct embedding searches over domain-specific knowledge ineffective.

To address this, IOAgent transforms each JSON summary into a natural language format for better alignment for the domain knowledge retrieval using language-based embedding similarity. This transformation is done by prompting an LLM with the code used to generate the summary, the JSON summary values, and a broader application context (total runtime, number of processes, and I/O size proportions from Darshan modules). An example of this transformation is shown in Figure \ref{pre_rag}, where the LLM provides a descriptive interpretation of the I/O size summary from the POSIX module. The natural language response aligns better with domain knowledge, improving the accuracy of embedding similarity searches. For instance, the JSON summary may contain: ``\textit{read\_histogram: \{`0-100': 1.0\}}", but the corresponding natural language would become: ``\textit{...the value of 1.0 in the 1 to 100 bin indicates that 100\% of the read operation fall within the 0 KB to 100 KB range}". The latter representation makes it much easier to match relevant studies in publications.

\subsubsection{Vector Database Creation}
The quality of the RAG database profoundly impacts the accuracy of subsequent diagnoses. However, to the best of our knowledge, there is no existing collection of text embeddings specifically targeting up-to-date HPC I/O performance diagnosis. Creating such a resource poses challenges, including the limits on how much data can be feasibly collected, embedded, and stored, which requires filtering and uncertainty about the best data sources. To address these issues and gather relevant, current knowledge on HPC I/O performance, we surveyed the past five years of research using the query `HPC I/O Performance' and `I/O Performance issue' in the ACM Digital Library and IEEE Xplore databases. From the top 50 results in each, we manually filtered for relevance, yielding 66 key works.

To process these works into a vector index for retrieval by IOAgent, we use LlamaIndex \cite{Liu_LlamaIndex_2022}, a popular open-source framework for vector index creation. LlamaIndex allows for configuring common hyperparameters such as the embedding model, chunk size, and overlap between chunks. We found retrieval accuracy and relevance to be consistent with reasonable variations in these settings, so we used the default configuration: a chunk size of 512, an overlap of 20, measured distance between embeddings via cosine similarity, and the \textit{text-embedding-3-large} model from OpenAI.


\begin{table*}[t]
    \centering
    \begin{tabularx}{\textwidth}{lX}
        \toprule
        \textbf{Label} & \textbf{Description} \\ \midrule
        High Metadata Load & The application spends a significant amount of time performing metadata operations (e.g., directory lookups, file system operations). \\
        Misaligned [Read\textbar Write] Requests & The application makes read or write requests that are not aligned with the file system's stripe boundaries. \\
        Random Access Patterns on [Read\textbar Write] & The application issues read or write requests in a random access pattern. \\
        Shared File Access & The application has multiple processes or ranks accessing the same file. \\
        Small [Read\textbar Write] I/O Requests & The application is making frequent read or write requests with a small number of bytes. \\
        Repetitive Data Access on Read & The application is making read requests to the same data repeatedly. \\
        Server Load Imbalance & The application issues a disproportionate amount of I/O traffic to some servers compared to others or does not properly utilize the available storage resources. \\
        Rank Load Imbalance & The application has MPI ranks issuing a disproportionate amount of I/O traffic compared to others. \\
        Multi-Process Without MPI & The application has multiple processes but does not leverage MPI. \\
        No Collective I/O on [Read\textbar Write] & The application does not perform collective I/O on read or write operations. \\
        Low-Level Library on [Read\textbar Write] & The application relies on a low-level library like STDIO for a significant amount of read or write operations outside of loading/reading configuration or output files. \\
        \bottomrule
    \end{tabularx}
    \caption{Table of I/O Issues and Descriptions}
    \label{tab:io_issues}
\end{table*}

\subsubsection{Vector Database Search}
With the vector index implemented as described, IOAgent conducts a vector search to match specific, related domain knowledge to each summary fragment as shown in Figure~\ref{ioagent}. Since in many cases the most useful information available in the vector index for a given summary fragment may not be the most pertinent and potential additional context provided by other highly ranked but not highest ranked matches may still be incredibly useful for an accurate diagnosis, IOAgent retrieves the top 15 closest matches from the index. However, since this enlarges the context significantly for any subsequent query that should analyze the content of these retrieved sources together, we implemented a \textit{self-reflection} step which uses a faster and cheaper language model (e.g., gpt-4o-mini) with less reasoning capability to rule out any sources which are found not to be relevant to the given summary fragment used as the original query over the index. This source filtering is run in parallel over all retrieved sources and rules out nearly half of the retrieved sources based on a more nuanced understanding of relevance than what is provided by the vector retriever. Following this parallel filtering step, IOAgent conducts its first true diagnosis of any potential I/O performance issues based on the information in the descriptive summary fragment generated by the LLM, as exemplified in Figure \ref{pre_rag}, and the filtered domain knowledge retrieved from the vector index. 

\subsection{Tree-based Merge}
Once IOAgent has generated an initial diagnosis for each summary category in the Darshan modules using the retrieved domain knowledge, these diagnoses—and their relevant references—must be merged into a single, comprehensive I/O performance diagnosis for the entire application.


Intuitively, one could merge all of the fragmented diagnoses and their source content via a single LLM query, as it is possible that all of this information could fit within the context window of some larger models. However, as later evaluations will show, the task of effectively and consistently merging more than two diagnosis summaries is beyond the capabilities of existing LLMs, even proprietary ones. This is primarily because regardless of the size of an LLM’s context window, the merging task itself requires diligent and precise reasoning. Effectively merging two diagnoses with their respective references involves removing redundant information, resolving contradictory details, and reflecting on both sets of provided sources to decide which should be retained in the new summary and where they should be cited. Additionally, adding any more summaries to the merging task beyond the minimal set of two introduces a more significant positional bias due to the increased cognitive load required in such a setting \cite{cogload,lostinmiddle}.

In light of the challenges associated with merging diagnosis summaries, IOAgent implements a pairwise merging strategy, in which two diagnosis fragments and their domain knowledge references are merged into a new, combined diagnosis and set of references. Since all pairs of diagnoses merged at the same level of the tree are independent of each other, all merging tasks for each level are conducted in parallel. The partial summaries are then further merged, effectively forming a tree structure, as shown in Figure~\ref{ioagent}. This tree-based merging plays a critical role in delivering concise and accurate diagnoses, as later evaluation results demonstrate.

%% file: sections/5_dataset.tex
\section{Benchmark Suite: TraceBench}
\label{sec:dataset}

Accurate evaluation of HPC I/O trace analysis tools presents a challenge due to the lack of publicly available trace datasets with labeled I/O performance issues. To address this, in line with our key contributions, we constructed the \textit{TraceBench} dataset, which includes a set of Darshan traces from three different sources. Each trace has been manually labeled based on a predefined set of labels covering common HPC I/O performance issues, as represented in Table \ref{tab:io_issues}.

The first data source consists of a set of Darshan logs generated by simple C source codes that intentionally introduce at least one of the I/O performance issues defined in the label set. We refer to this set as \textit{Simple-Bench}. The second data source consists of a set of Darshan logs generated by the IO500 benchmark \cite{io500}, where each configuration is designed to induce one or more I/O performance issues defined by the labels. The final data source primarily comprises real application traces, all collected on production HPC systems.

\textbf{1) Simple-Bench (SB).} The Simple-Bench set consists of 10 labeled Darshan logs, which were generated using 10 rudimentary scripts written in C to target at least one specific I/O performance issue category from Table \ref{tab:io_issues}. However, as shown by the sample representation across different issue categories for each dataset in Table \ref{tab:data_rep}, some traces from the Simple-Bench source contain more than one issue type. 
Despite this, the simplicity of these scripts results in Darshan trace logs that are small in size, with very low aggregate I/O volume and highly uniform behavior. As a result, these traces should be the easiest to diagnose accurately among all tools.

\textbf{2) IO500.}
The second set of labeled Darshan trace logs was collected using the IO500 benchmark, a synthetic performance benchmark tool for HPC systems that simulates a broad set of commonly observed I/O patterns in HPC applications~\cite{io500}. IO500 consists of many configurable workloads, which can be tuned to simulate various sub-optimal I/O patterns. For example, IO500’s ior-easy workload, which conducts intense sequential read and write phases, can be tuned to use 8k transfer sizes issued through independent POSIX operations across multiple ranks, resulting in dominant small I/O behavior that does not effectively leverage higher-level libraries such as MPI-IO for multi-rank I/O. 
In total, 21 traces were collected from 21 unique configurations of IO500, and as shown in Table \ref{tab:data_rep}, a significant number of traces exhibit multiple overlapping issues.

\textbf{3) Real Applications (RA).}
The final set of labeled Darshan trace logs was generated by running real application workloads on large-scale production HPC systems. This collection consists of nine samples, each originating from a unique application, except for two samples representing recollected traces for the E2E and OpenPMD applications. In these cases, the original trace for each application possessed a significant performance issue, and the recollected traces had their primary issues resolved.

\begin{table}[ht!]
    \centering
    \begin{tabular}{l|c|c|c|c}
        \toprule
        \textbf{Labeled Issue} & \textbf{SB} & \textbf{IO500} & \textbf{RA} & \textbf{Total} \\ \hline \hline
        High Metadata Load & 1 & 2 & 2 & 5 \\ \hline
        Misaligned Read requests & 2 & 10 & 4 & 16 \\ \hline
        Misaligned Write requests & 2 & 10 & 6 & 18 \\ \hline
        Random Access Patterns on Write & 0 & 5 & 2 & 7 \\ \hline
        Random Access Patterns on Read & 0 & 5 & 2 & 7 \\ \hline
        Shared File Access & 1 & 14 & 4 & 19 \\ \hline
        Small Read I/O Requests & 2 & 10 & 5 & 17 \\ \hline
        Small Write I/O Requests & 2 & 10 & 6 & 18 \\ \hline
        Repetitive Data Access on Read & 1 & 0 & 0 & 1 \\ \hline
        Server Load Imbalance & 7 & 15 & 2 & 24 \\ \hline
        Rank Load Imbalance & 1 & 0 & 1 & 2 \\ \hline
        Multi-Process W/O MPI & 0 & 13 & 0 & 13 \\ \hline
        No Collective I/O on Read & 6 & 8 & 4 & 18 \\ \hline
        No Collective I/O on Write & 5 & 8 & 2 & 15 \\ \hline
        Low-Level Library on Read & 1 & 0 & 0 & 1 \\ \hline
        Low-Level Library on Write & 1 & 0 & 0 & 1 \\ \hline
    \end{tabular}
    \caption{Summary of traces and labeled issues.}
    \label{tab:data_rep}
\end{table}

Table~\ref{tab:data_rep} summarizes all the I/O issues included in \texttt{TraceBench} and how different sources, such as SB (Single-Bench), IO500, RA (Real-Application), contribute to these issues. There are 182 issues reported in 40+ traces.

%% file: sections/6_experiments.tex
\section{Evaluation}
\label{sec:Eval}
\textbf{Summary.}
We conducted comprehensive evaluations of IOAgent using the \texttt{TraceBench} test suite. The diagnosis results were compared to those from the LLM-based diagnosis tool ION~\cite{egersdoerfer2024ion} and the expert-guided I/O performance diagnosis tool Drishti~\cite{drishti}. Additionally, we evaluated examples of advanced user interactions enabled by IOAgent’s LLM-centric design and validated our tree-based merge design, as outlined in Section \ref{sec:design}.


\subsection{Evaluation Metrics}

To evaluate IOAgent, an \textit{accuracy} metric, which measures how well the diagnosis matches the ground truth, is certainly the most important criterion to consider. With this objective in mind, we expect IOAgent to perform on par with state-of-the-art expert-based diagnosis tools, such as Drishti. However, accuracy should not be the only critical metric evaluated. Since IOAgent is intended to assist domain scientists, the readability and understandability of the information provided in the diagnosis also become essential for users at any level of familiarity with HPC I/O. As an assistant, IOAgent should also be assessed based on how useful the information is.
Following practices from the NLP community for evaluating agents~\cite{liu2023gevalnlgevaluationusing, RagasDocumentation, llmasajudge}, we propose the following three metrics for our evaluations:
\begin{itemize}
\item \textit{Accuracy}: evaluate how accurately the ground truth labels are diagnosed by each tool.
\item \textit{Utility}: evaluate how useful the information provided in each diagnosis is for understanding the overall I/O behavior of the application, identifying performance issues, and determining how to address each noted issue (regardless of the factuality of such statements).
\item \textit{Interpretability}: evaluate how readable and understandable the provided information is for users at any level of familiarity with HPC I/O.
\end{itemize}

\begin{table*}[ht!]
\centering
\caption{Performance Results for Diagnosis Tools on TraceBench Subsets}
\label{tab:performance_metrics}
\begin{tabular}{
    l l
    S[table-format=1.3] 
    S[table-format=1.3] 
    S[table-format=1.3] 
    S[table-format=1.3] 
    }
\toprule
\textbf{Metric} & \textbf{Diagnosis Tool} & \textbf{Simple-Bench} & \textbf{IO500} & \textbf{Real-Applications} & \textbf{Overall} \\
\midrule
\multirow{4}{*}{Accuracy} & Drishti               & 0.398 & 0.480 & 0.472 & 0.459 \\
                          & ION                   & 0.343 & 0.381 & 0.417 & 0.380 \\
                          & IOAgent-gpt-4o        & \bfseries 0.630 & \bfseries 0.655 & \bfseries 0.620 & \bfseries 0.641 \\
                          & IOAgent-llama-3.1-70B & 0.620 & 0.488 & 0.463 & 0.513 \\
\midrule
\multirow{4}{*}{Utility}  & Drishti               & 0.426 & 0.417 & 0.491 & 0.436 \\
                          & ION                   & 0.352 & 0.401 & 0.380 & 0.385 \\
                          & IOAgent-gpt-4o        & 0.565 & \bfseries 0.615 & \bfseries 0.639 & \bfseries 0.609 \\
                          & IOAgent-llama-3.1-70B & \bfseries 0.694 & 0.587 & 0.565 & 0.607 \\
\midrule
\multirow{4}{*}{Interpretability} & Drishti        & 0.417 & 0.452 & 0.463 & 0.447 \\
                                  & ION            & 0.343 & 0.417 & 0.352 & 0.385 \\
                                  & IOAgent-gpt-4o & 0.546 & \bfseries 0.659 & \bfseries 0.713 & \bfseries 0.645 \\
                                  & IOAgent-llama-3.1-70B & \bfseries 0.694 & 0.484 & 0.472 & 0.530 \\
\midrule
\multirow{4}{*}{Average}            & Drishti        & 0.414 & 0.450 & 0.475 & 0.447 \\
                                  & ION            & 0.346 & 0.399 & 0.383 & 0.383 \\
                                  & IOAgent-gpt-4o & 0.580 & \bfseries 0.643 & \bfseries 0.657 & \bfseries 0.632 \\
                                  & IOAgent-llama-3.1-70B & \bfseries 0.670 & 0.520 & 0.500 & 0.550 \\
\bottomrule
\end{tabular}
\vspace{-1em}
\end{table*}

\subsection{LLM-based Rating System}
With the metrics defined, the challenge shifts to providing a quantitative judgment of each metric. Among them, \textit{accuracy} is relatively easy as we can easily count the number of matched and mismatched issues. But \textit{Utility} and \textit{Interpretability} are rather subjective to individuals and their experience level with HPC I/O. In this study, we use LLM as the judge and follow the common practice of using a ranking-based system to compare different solutions.

The intuition behind using a capable LLM to rate the diagnostic results is simple. Earlier results in Figure~\ref{llm-example} have shown a qualified LLM, such as GPT-4o, is very capable of understanding high-level I/O relevant concepts (with misunderstanding in many details of course), which highly emulates our target users: domain scientists. We believe that letting capable LLM serve as the judge avoids personal bias, and more importantly, brings in the perspectives of domain users. 

Even with a capable LLM, quantitatively providing a score on \textit{Utility} or \textit{Interpretability} is still not feasible. Instead, we use an anonymized rating system to compare different solutions and rank them. The ranking is done by the LLM itself by formatting a prompt with all diagnosis outputs from different tools, a description of the specified evaluation criteria, and a description of how the ranking output should be formatted. The LLM, GPT-4o in this case, will rank the diagnosis outputs of different tools on a scale of 1 to 4, with 1 being the best and 4 being the worst.  

\begin{figure}[ht!]
    \centering
\includegraphics[width=\linewidth]{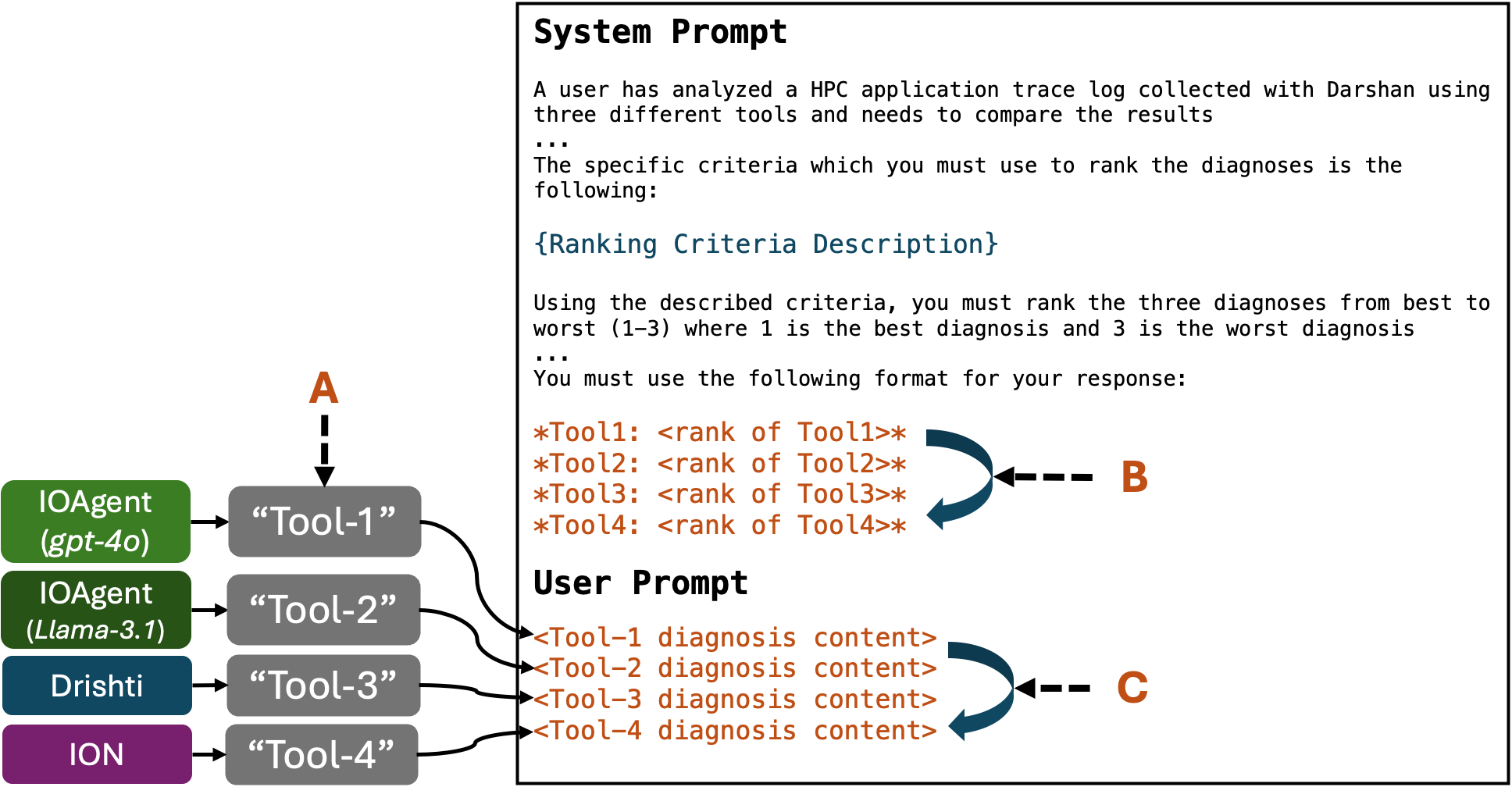}
    \caption{A, B and C denote three augmentations applied.}
    \label{ranking_setup}
\end{figure}

Due to the known potential for \textit{positional bias} in LLM-based content ranking \cite{lostinmiddle, shi2024judgingjudgessystematicinvestigation} we further augment the prompt in three key ways to ensure a fair and reliable ranking result. The first augmentation is to anonymize the names of the diagnosis tools to eliminate any potential bias towards or against specific analysis tools. The second augmentation is to rotate the rank assignment order imposed by the response formatting portion of the prompt. The final augmentation is to rotate the order in which the content of each diagnosis appears in the prompt. All these measures are intended to eliminate positional bias for each diagnosis in the prompt context. These augmentations are outlined in Figure~\ref{ranking_setup}, denoted by the letters A, B, and C. Additionally, for each sample, we rank the diagnoses four times, ensuring that every ranking prompt permutation appears at least once, further reducing potential bias.


\subsection{Quantitative Score Calculation}
Once the diagnoses have been ranked, we calculate the aggregated score using the following approach: For each trace log $L$, the diagnosis result from each diagnosis tool $T$ will be evaluated three times on diagnosis criterion $ C \in \{ \text{Accuracy}, \text{Utility}, \text{Interpretability} \} $ and assigned a $Rank \in [1,4]$. The score for each sample is calculated as $S_{T,C,L} = (4 - \text{Rank}_{T, C, L})$. A lower numerical rank (e.g. Rank 1) indicates a higher or better score.

The total score over all samples for a given data source $D \in \{ \text{Simple-Bench}, \text{IO500}, \text{Real-Applications} \}$ is therefore defined as the sum of $S_{T,C,L}$: 
\begin{equation}
S_{T,C,D} = \sum_{L \in D} S_{T, C, L}
\label{directory_sum}
\end{equation}
We then normalize the score into a value between 0 and 1 by dividing it by the maximal score one can get: 
\begin{equation}
NS_{T, C, D} = \frac{S_{T, C, D}}{(4-1) \cdot |D|}
\label{directory_sum}
\end{equation}
Here $(4-1)$ is the highest score for each trace in $D$. 
Based on $NS_{T, C, D}$, we can define the average score of each tool across all three metrics over all data sources to show how the tool works across metrics. 
It is simply an average of $NS_{T,C,D}$ over all three metrics. 
Similarly, we also define the average score of each tool across all data sources on each metric to show how the tool works across different data sources. 

Note that, to minimize hallucination by the LLM during ranking, we require the model to verify the reasoning behind the assigned positions by including a rank assignment explanation as part of the prompt. This approach provides significantly more insight into the assigned scores, as demonstrated by the following example of an explanation provided by the ranking LLM:

\begin{quotation}
Tool-2 and Tool-3 both provided comprehensive diagnoses that accurately identified all five I/O performance issues: small read and write I/O requests, misaligned read and write requests, and the use of multiple processes without MPI. Tool-2 provided detailed recommendations and emphasized the need to align with the file system boundaries, making it a strong contender for the top rank.
\end{quotation}

\subsection{Ranking Results}
Table~\ref{tab:performance_metrics} summarizes all the results. We present the results for each metric (i.e., accuracy, utility, and interpretability) as well as the average across all metrics in the rows. The results for each type of job trace in \texttt{TraceBench}, along with the overall average, are shown in the columns. Each individual row represents one diagnosis tool, including the naive LLM-based tool (ION) using gpt-4o (\textit{gpt-4o-2024-05-13}) as its backbone and the state-of-the-art heuristic tool (Drishti). For IOAgent, we include two instances, labeled “IOAgent-*”: one uses the proprietary gpt-4o (\textit{gpt-4o-2024-05-13}) model from OpenAI, and the other uses the open-source \textit{Llama-3.1-70B-Instruct} model from Meta. By including results from both instances, we want to evaluate that if IOAgent relies on specific proprietary models.

\begin{figure}[t!]
    \centering
\includegraphics[width=\linewidth]{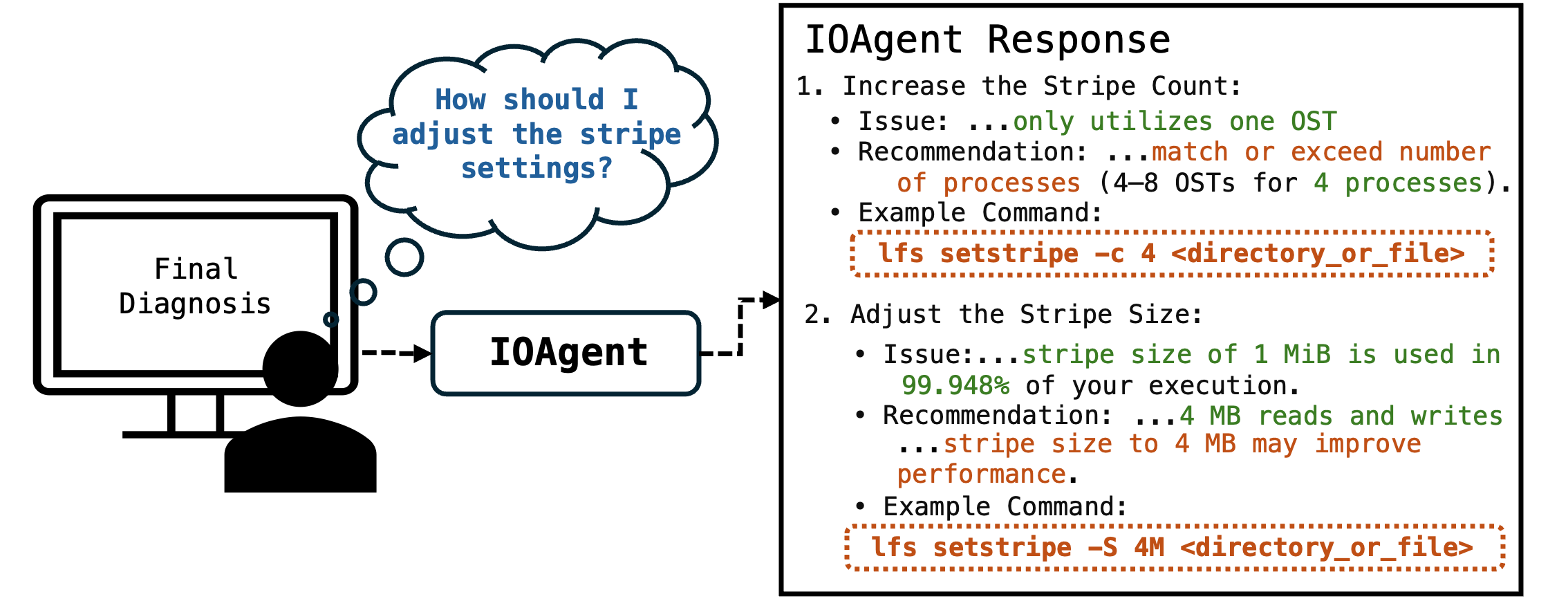}
    \caption{Post-diagnosis example between user and IOAgent.}
    \label{user_interaction}
    \vspace{-1em}
\end{figure}

\begin{figure*}[t!]
    \centering
\includegraphics[width=\textwidth]{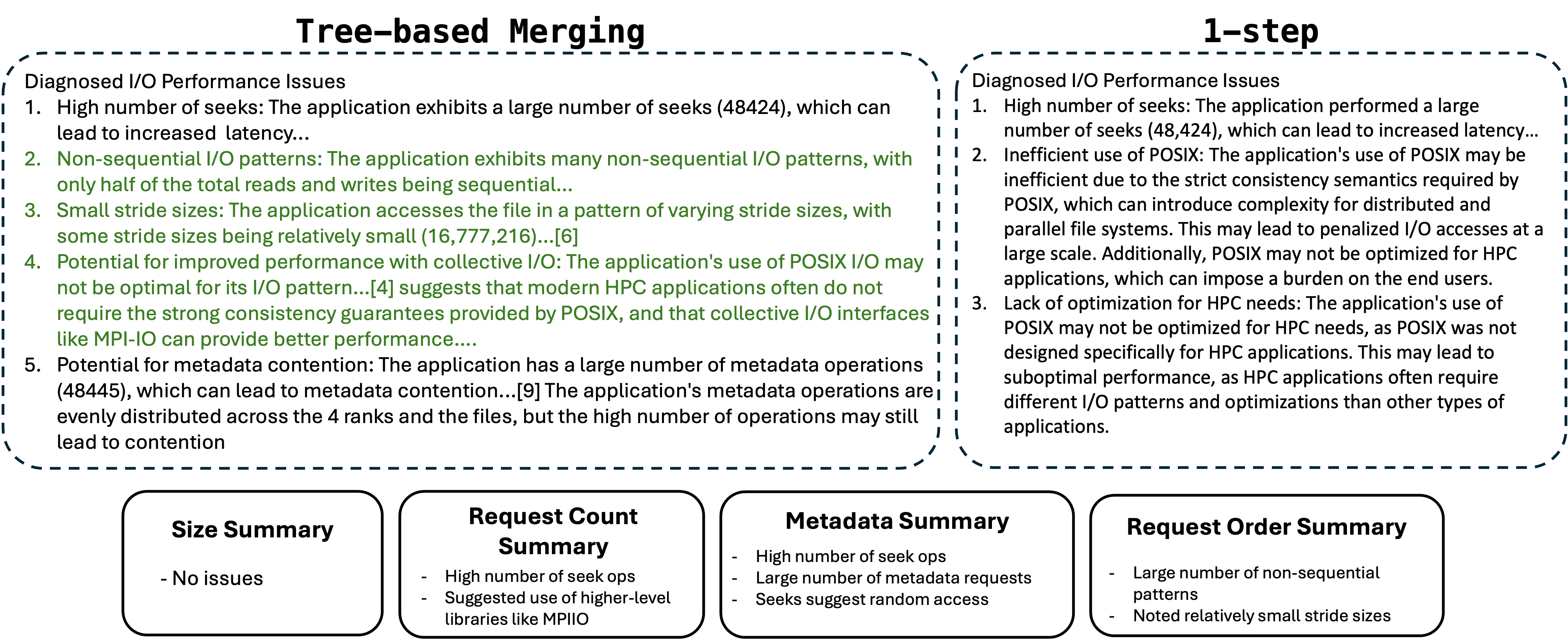}
    \caption{Comparison of output using pair-wise vs 1-step merge between 4 diagnosis summaries}
    \label{ablation_tree}
    \vspace{-1em}
\end{figure*}

The results in Table~\ref{tab:performance_metrics} clearly show that both IOAgent tools perform exceptionally well across all metrics and job traces. As expected, IOAgent-gpt-4o performs particularly well, given that it uses a frontier-level LLM; however, it is noteworthy that IOAgent-llama-3.1-70B also performs strongly. Its results are close to those of the gpt-4o model and surpass those of the state-of-the-art heuristic tool (Drishti) and the naive LLM-based tool (ION). Interestingly, Llama IOAgent seems to excel in more primitive cases like Simple-Bench. This may be due to IOAgent-gpt-4o providing too many details in such basic cases, making its output less useful or understandable from the user's perspective. These results confirm that our carefully designed IOAgent workflow makes LLMs a stable and practical tool for assisting scientists in understanding their applications’ I/O issues.


\subsection{Continued User Interaction}

A primary feature enabled by the LLM-centric design of IOAgent is the ability for users to gain a deeper understanding and receive application-specific recommendations with implementation guidance through continued chat-based interaction. To evaluate this feature, we present a real interaction case summarized in Figure \ref{user_interaction}. In this example, IOAgent analyzed a log from the IO500 \texttt{TraceBench} subset, which performed a significant number of 4MB reads and writes but used the default Lustre stripe width and stripe size parameters of 1 and 1MB, respectively. The final diagnosis highlighted these potentially suboptimal stripe settings.


Following the diagnosis, the user simply prompted IOAgent about how to fix such an issue (highlighted in blue). In this case, IOAgent effectively utilized the context of the diagnosis and its referenced sources to provide detailed assistance, offering explanations of recommended actions and code samples, such as \texttt{lfs setstripe -S 4M} (highlighted in orange). These responses were not only accurate but also tailored to the specific issue identified by the application (highlighted in green). Such cases demonstrate IOAgent’s unique potential to effectively assist domain scientists at scale.


\subsection{Why Tree-based Merge?}

The \textit{tree-based merge} mechanism introduces both additional time and monetary overhead, as merging is done in pairs rather than all at once through a single prompt. In this evaluation, we aim to demonstrate the necessity of this overhead. Specifically, we present a comparison of using the tree-based merge versus not using it, as shown in Figure \ref{ablation_tree}. For this benchmark, we used a less capable open-source \textit{Llama-3-70B} model to illustrate how direct merging can be problematic. Due to space constraints, we use the example of merging just four summary diagnoses: \textit{Size}, \textit{Request Count}, \textit{Metadata}, and \textit{Request Order}. At the bottom of Fig.~\ref{ablation_tree}, we briefly list the issues identified with each category. We then compare the merged results of our tree-based approach with the naive one-step solution, using the same prompts.



The results of the 1-step merge show that important information regarding the non-sequential I/O patterns and insight into the stride sizes as well as the specific recommended use of higher level parallel I/O libraries (MPI-IO) are all lost, along with their respective domain reference sources. In contrast, the tree-based merge successfully maintains the key points of each individual summary diagnosis as well as the useful domain reference sources pertaining to each. 
Note that, this is just a simpler case where only four summaries need to be merged. In IOAgent, we typically need to deal with 13 summary diagnoses, which is extremely challenging even for the latest gpt-4o model based on our experiments. 

%% file: sections/7_conclusion.tex
\section{Conclusion and Future Work}
\label{sec:conclude}
In this study, we propose and implement IOAgent, a model-agnostic LLM-based framework for HPC I/O trace log analysis which fills the existing gaps faced by non-augmented language models while analyzing logs.
IOAgent implements a module-based log pre-processor, a vector index for domain information retrieval, and a tree-based summarization strategy to enable accurate, context-aware, and user-friendly I/O performance analysis achieving results on-par with or superior to existing, expert-level tools as evaluated on multiple criteria. Leveraging the conversational strengths of language models further democratizes access to interactive and accurate I/O performance diagnoses. Our future work includes the expansion of IOAgent's existing capabilities to provide in-depth user interaction capabilities, such as building the continually advancing reasoning capabilities of LLMs and the integration of continual analysis improvements through interaction.